\documentclass[]{llncs}

\usepackage{theorem}
\theorembodyfont{\rmfamily}

\theoremstyle{plain}

\usepackage{prooftree,hyperref,amsmath,amssymb,upgreek,breakurl,textcomp,centernot,graphicx}
\usepackage[usenames,dvipsnames]{color}
\usepackage{tikz}
\usetikzlibrary{matrix,arrows}

\newcommand\evolve[1]{\mathop\square #1} 

\newcommand\astep[1]{\stackrel{#1}{\longrightarrow}} 

\newcommand\gramto{\mathrel{::=}} 

\newcommand\dom[1]{\mathrm{dom}(#1)}

\newcommand\blue[1]{\textcolor{blue}{#1}}
\newcommand\step{\mathrel{\blue\to}}
\newcommand\steps{\mathrel{\step\cdots\step}}

\newcommand\kleene[1]{{#1}{}^{*}\mkern-1.0mu}

\newcommand\infer[2]{\begin{prooftree} #1 \justifies #2  \end{prooftree}}
\newcommand\mpair[2]{#1 \mathrel{\blue ;} #2}
\newcommand\mconf[1]{\blue{\langle} #1 \blue{\rangle}}

\newcommand\knode[1]{\mathop{\normalfont \texttt{cont}} #1}

\newcommand\nullpointer{{\normalfont\texttt{null}}}
\newcommand\mklist[1]{[#1]} 
\newcommand\empstring{\varepsilon} 
\newcommand\epsexp{\mathtt{\varepsilon}} 
\newcommand\emplist{\mklist{\,}} 
\newcommand\cons[2]{#1\mkern3mu{:\mkern-1.5mu:}\mkern3mu#2} 
\newcommand\conc[2]{#1@#2} 

\newcommand\pwpi[2]{\mconf{\mpair{#1}{#2}}}

\newcommand\empset{\emptyset}

\renewcommand\mpair[2]{#1 \mkern4mu\blue{;}\mkern4mu #2}

\newcommand\pwfpi[3]{\cons{\pwpi{#1}{#2}}{#3}}

\renewcommand\cons[2]{#1 :\mkern-2mu: #2}
\renewcommand\conc[2]{#1 \mathrel\cdot #2}

\numberwithin{equation}{section}
\numberwithin{figure}{section}

\newcommand\seqcomp{\cdot} 

\title{Static Analysis for Regular Expression Denial-of-Service Attacks}

\author{James Kirrage \qquad Asiri Rathnayake \qquad Hayo Thielecke\\
\institute{University of Birmingham, UK}
}

\begin{document}

\maketitle

\begin{abstract}
Regular expressions are a concise yet expressive language for expressing patterns. For instance, in networked software, they are used for input validation and intrusion detection. Yet some widely deployed regular expression matchers based on backtracking are themselves vulnerable to denial-of-service attacks, since their runtime can be exponential for certain input strings. This paper presents a static analysis for detecting such vulnerable regular expressions. The running time of the analysis compares favourably with tools based on fuzzing, that is, randomly generating inputs and measuring how long matching them takes. Unlike fuzzers, the analysis pinpoints the source of the vulnerability and generates possible malicious inputs for programmers to use in security testing. Moreover, the analysis has a firm theoretical foundation in abstract machines. Testing the analysis on two large repositories of regular expressions shows that the analysis is able to find significant numbers of vulnerable regular expressions in a matter of seconds.
\end{abstract}

\section{Introduction}

Regular expression matching is a ubiquitous technique for reading and validating input, particularly in web software. 
While pattern matchers are among the standard techniques for defending against malicious input, they are themselves vulnerable. The root cause of the vulnerability is that
widely deployed regular expression matchers, like the one in the Java libraries, are based on \emph{backtracking} algorithms, rather than the construction of a  Deterministic Finite Automaton (DFA), as used for lexers in compiler construction~\cite{hopcroftullman,2007_dragonbook}. One reason for relying on backtracking rather than a DFA construction is to support a
  more expressive pattern specification language commonly referred to as ``regexes". Constructs such as back-references supported by such regex languages go beyond regular and even context-free languages and are known to be computationally expensive~\cite{1990_aho}. However, even if restricted to purely regular constructs, backtracking matchers may have a running time that is exponential in the size of the input~\cite{2007_regex_cox}, potentially causing a regular expression denial-of-service (ReDoS) attack~\cite{2012_redos_owasp}. It is this potentially exponential runtime on pure regular expressions (without backreferences) that we are concerned about in this paper. Part of our motivation is that, for purely regular expressions, the attack could be defended against by avoiding backtracking matchers and using more efficient techniques~\cite{2009_regex_cox,1968_thompson} instead. 
  
For a minimalistic example~\cite{2007_regex_cox}, consider matching the regular expression \texttt{a**} against the input string \texttt{a}\ldots\texttt{a}\;\texttt b, with $n$ repetitions of \texttt a. A backtracking matcher  takes an exponential time~\cite{2007_regex_cox} in $n$ when trying to find a match; all matching attempts fail in the end due to the trailing \texttt b.
For such vulnerable regular expressions, an attacker can craft an input of moderate size which causes the matcher to take so long that for all practical purposes the matcher fails to terminate, leading to a denial-of-service attack. Here we assume that the regular expression itself cannot be manipulated by the attacker but that 
it is matched against a string that is user-malleable.

While the regular expression \texttt{a**} as above is contrived, one of the questions we set out to answer is how prevalent such vulnerable expressions are in the real world. As finding vulnerabilities manually in code is time consuming and error-prone, there is growing interest in automated tools for static analysis for security~\cite{livshits2005finding,chess2004static}, motivating us to design an analysis for ReDoS.

Educating and warning programmers is crucial to defending against attacks on software. The standard coverage of regular expressions in the computer science curriculum, covering DFAs in courses on computability~\cite{hopcroftullman} or compiler construction~\cite{2007_dragonbook}, is not necessarily sufficient to raise awareness about the possibility of ReDoS. Our analysis constructs a series of attack strings, so that developers can confirm the exponential runtime for themselves.


This paper makes the following contributions:
\begin{enumerate}
\item We present an efficient static analysis for DoS on pure regular expressions.
\item The design of the tool has a firm theoretical foundation based on abstract machines~\cite{2011_regexpsos} and derivatives~\cite{1964_brzozowski} for regular expressions.
\item We report finding vulnerable regular expressions in the wild.
\end{enumerate}

In Section~\ref{secmachine}, we describe backtracking regular expression matchers as abstract machines, so that we have a precise model of what it means for a matching attempt to take an exponential number of steps. We  build on the abstract machine in designing our static analysis in Section~\ref{staticanalysis}, which we have implemented in OCaml as described in Section~\ref{implementation}. Experimental results in testing the analysis on two large corpora of regular expressions  are reported in  Section~\ref{results}. Finally, Section~\ref{conclusions} concludes with a discussion of related work and directions of further research.
The code of the tool and data sets are available at this URL: \\ \url{http://www.cs.bham.ac.uk/~hxt/research/rxxr.shtml}

\section{Regular expression matching by backtracking}
\label{secmachine}

This and the next section present the theoretical basis for our analysis. Readers primarily interested in the results may wish to skim them.

We start with the following minimal  syntax for regular expressions:
\[
\begin{array}{rcl@{\hspace{4em}}l}
e &\gramto& e_{1} \mid e_{2} & \mbox{Alternation}
 \\[1ex]
&& \kleene e &\mbox{Kleene star}
\\[1ex]
&& e_{1} \cdot e_{2} &\mbox{Concatenation}
\\[1ex]
&& \texttt a  &\mbox{Constant, where \texttt a is an input symbol}
\end{array}
\]
The $\cdot$ in concatenation $e_{1} \cdot e_{2}$ is usually omitted, except when it is useful for emphasis, as in a syntax tree.
Following the usual parser construction methods~\cite{2007_dragonbook}, we can define a parser which is capable of transforming (parsing) a given regular expression into an AST (abstract syntax tree) which complies with the above grammar. As an example, the AST constructed by such a parser for the regular expression $\kleene{(a \mid b)}c$ can be visualized in the following manner:

\vspace{4mm}
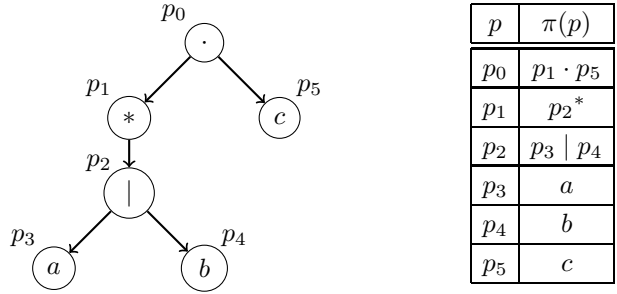
\begin{minipage}{0.4\textwidth}\centering
  \begin{tikzpicture}
    \path (-0.4, 0.4)       node[rectangle] (b0) {$p_{0}$};
    \path (0, 0)            node[draw, circle] (p0) {$\seqcomp$};

    \path (-1.4, -0.6)      node[rectangle] (b1) {$p_{1}$};
    \path (-1, -1)          node[draw, circle] (p1) {$*$};

    \path (-1.4, -1.6)      node[rectangle] (b2) {$p_{2}$};
    \path (-1, -2)          node[draw, circle] (p2) {$\mid$};

    \path (-2.4, -2.6)      node[rectangle] (b3) {$p_{3}$};
    \path (-2, -3)          node[draw, circle] (p3) {$a$};

    \path (0.4, -2.6)      node[rectangle] (b4) {$p_{4}$};
    \path (0, -3)           node[draw, circle] (p4) {$b$};

    \path (1.4, -0.6)       node[rectangle] (b5) {$p_{5}$};
    \path (1, -1)           node[draw, circle] (p5) {$c$};

    \draw[->, thick](p0) -- (p1);
    \draw[->, thick](p0) -- (p5);
    \draw[->, thick](p1) -- (p2);
    \draw[->, thick](p2) -- (p3);
    \draw[->, thick](p2) -- (p4);
  \end{tikzpicture}
\end{minipage}
\begin{minipage}[!t]{0.4\textwidth}\centering
\def\arraystretch{1.2}
\setlength{\tabcolsep}{1ex}
  \begin{tabular}{|c|c|}
    \hline
    $p$ & $\pi(p)$\\
    \hline\hline
    $p_{0}$ & $p_{1} \seqcomp p_{5}$\\
    \hline
    $p_{1}$ & $\kleene{p_{2}}$\\
    \hline
    $p_{2}$ & $p_3 \mid p_4$\\
    \hline
    $p_{3}$ & $a$\\
    \hline
    $p_{4}$ & $b$\\
    \hline
    $p_{5}$ & $c$\\
    \hline
  \end{tabular}
\end{minipage}

\vspace{4mm}
\noindent
Notice that we have employed a pointer notation to illustrate the AST structure; this is quite natural given that in most programming languages, such an AST would be defined using a similar pointer-based structure definition. Each node of this AST corresponds to a unique sub-expression of the original regular expression, the relationships among these nodes are given on the table to the right. We have used the notation $\pi(p)$ to signify the dereferencing of the pointer $p$ with respect to the heap $\pi$ in which the above AST is constructed. A formal definition of $\pi$ was avoided in order to keep the notational clutter to a minimum, interested readers may refer~\cite{2011_regexpsos} for a more precise definition of $\pi$.

Having parsed the regular expression into an AST, the next step is to construct an NFA structure that allows us to define a backtracking pattern matcher. While there are several standard NFA construction techniques~\cite{2007_dragonbook}, we opt for a slightly different construction which greatly simplifies the rest of the discussion. The idea is to associate a continuation pointer $\knode$ with each of the nodes in the AST such that $\knode$ points to the \textit{following} (continuation) expression for each of the sub-expressions in the AST. In other words, $\knode$ identifies the ``next sub-expression" which must be matched after matching the given sub-expression. More formally, $\knode$ is defined as follows:
\begin{definition} \label{defknode}
Let $\knode$ be a function
\[
\knode{} : \dom{\pi} \to (\dom{\pi} \cup \{ \nullpointer \})
\]
Such that,
\begin{itemize}
\item
If $\pi(p) = (p_{1} \mid p_{2})$, then $\knode{p_{1}} = \knode{p}$ and $\knode{p_{2}} = \knode{p}$
\item
If $\pi(p) = (p_{1} \seqcomp p_{2})$, then $\knode{p_{1}} = p_{2}$ and $\knode{p_{2}} = \knode{p}$
\item
If $\pi(p) = \kleene{(p_{1})}$, then $\knode{p_{1}} = p$
\item
$\knode{p_{0}} = \nullpointer$, where $p_{0}$ is the pointer to the root of the AST.
\end{itemize}
\end{definition}
The following example illustrates the NFA constructed this way for the regular expression $\kleene{(a \mid b)}c$:

\vspace{4mm}
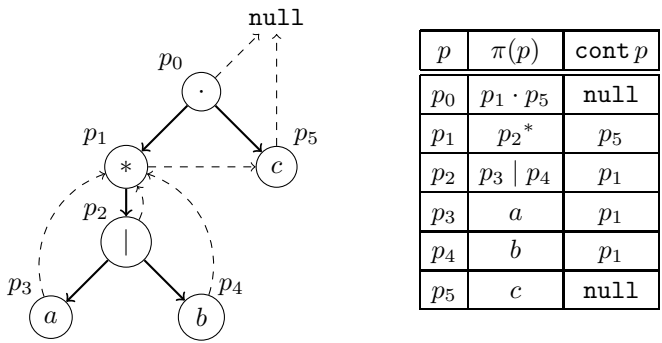
\begin{minipage}{0.4\textwidth}\centering
  \begin{tikzpicture}
    \path (1, 1)            node[rectangle] (null) {$\nullpointer$};

    \path (-0.4, 0.4)       node[rectangle] (b0) {$p_{0}$};
    \path (0, 0)            node[draw, circle] (p0) {$\seqcomp$};

    \path (-1.4, -0.6)      node[rectangle] (b1) {$p_{1}$};
    \path (-1, -1)          node[draw, circle] (p1) {$*$};

    \path (-1.4, -1.6)      node[rectangle] (b2) {$p_{2}$};
    \path (-1, -2)          node[draw, circle] (p2) {$\mid$};

    \path (-2.4, -2.6)      node[rectangle] (b3) {$p_{3}$};
    \path (-2, -3)          node[draw, circle] (p3) {$a$};

    \path (0.4, -2.6)      node[rectangle] (b4) {$p_{4}$};
    \path (0, -3)           node[draw, circle] (p4) {$b$};

    \path (1.4, -0.6)       node[rectangle] (b5) {$p_{5}$};
    \path (1, -1)           node[draw, circle] (p5) {$c$};

    \draw[->, thick](p0) -- (p1);
    \draw[->, thick](p0) -- (p5);
    \draw[->, dashed](p0) -- (null);
    \draw[->, dashed](p5) -- (null);

    \draw[->, thick](p1) -- (p2);
    \draw[->, dashed](p1) -- (p5);

    \draw[->, thick](p2) -- (p3);
    \draw[->, thick](p2) -- (p4);
    \draw[->] (p2) edge[bend right=30, dashed] (p1);
    \draw[->] (p3) edge[bend left=45, dashed] (p1);
    \draw[->] (p4) edge[bend right=45, dashed] (p1);
  \end{tikzpicture}
\end{minipage}
\begin{minipage}[!t]{0.4\textwidth}\centering
\def\arraystretch{1.2}
\setlength{\tabcolsep}{1ex}
  \begin{tabular}{|c|c|c|}
    \hline
    $p$ & $\pi(p)$ & $\knode{p}$\\
    \hline\hline
    $p_{0}$ & $p_{1} \seqcomp p_{5}$ & \nullpointer\\
    \hline
    $p_{1}$ & $\kleene{p_{2}}$ & $p_5$\\
    \hline
    $p_{2}$ & $p_3 \mid p_4$ & $p_1$\\
    \hline
    $p_{3}$ & $a$ & $p_1$\\
    \hline
    $p_{4}$ & $b$ & $p_1$\\
    \hline
    $p_{5}$ & $c$ & $\nullpointer$\\
    \hline
  \end{tabular}
\end{minipage}

\vspace{4mm}
\noindent
Here the dashed arrows identify the $\knode$ pointer for each of the AST nodes. Readers familiar with  Thompson's construction~\cite{1968_thompson,2007_dragonbook} will realize that the resulting NFA is a slightly pessimized version of that resulting from  Thompson's algorithm. The reason for this pessimization is purely of presentational nature; it helps to visualize the NFA as an AST with an overlay of a $\knode$ pointer mesh so that the structure of the original regular expression is still available in the AST portion. Furthermore, this presentation allows the definitions and proofs to be presented in an inductive fashion with respect to the structure of the expressions.

With the NFA defined, we present a simple non-deterministic regular expression matcher in the form of an abstract-machine called the PW$\pi$ machine:
\begin{definition}\label{defmatchmach}
A configuration of the PW$\pi$ machine consists of two components:
\[
\pwpi{p}{w}
\]
The $p$ component represents the current sub-expression (similar to a code pointer) while $w$ corresponds to the rest of the input string that remains to be matched. The transitions of this machine are as follows:
\begin{eqnarray*}
\pwpi{p}{w} &\step &\pwpi{p_1}{w} \textrm{ if } \pi(p) = (p_1 \mid p_2)\\
\pwpi{p}{w} &\step  &\pwpi{p_2}{w} \textrm{ if } \pi(p) = (p_1 \mid p_2)\\
\pwpi{p}{w} &\step&\pwpi{q}{w} \textrm{ if } \pi(p) = \kleene{p_1} \land \knode{p} = q\\
\pwpi{p}{w} &\step &\pwpi{p_1}{w} \textrm{ if } \pi(p) = \kleene{p_1}\\
\pwpi{p}{w} &\step &\pwpi{p_1}{w} \textrm{ if } \pi(p) = (p_1 \cdot p_2)\\
\pwpi{p}{aw} &\step &\pwpi{q}{w} \textrm{ if } \pi(p) = a \land \knode{p} = q\\
\pwpi{p}{w} &\step &\pwpi{q}{w} \textrm{ if } \pi(p) = \epsexp \land \knode{p} = q
\end{eqnarray*}
The initial state of the PW$\pi$ machine is $\pwpi{p_0}{w}$, where $p_0$ is the root of the AST corresponding to the input expression and $w$ is the input string. The machine may terminate in the state $\pwpi{\nullpointer}{w''}$ where it has matched the original regular expression against some prefix $w'$ of the original input string $w$ such that $w = w'w''$. Apart from the successful termination, the machine may also terminate if it enters into a configuration where none of the above transitions apply.
\end{definition}
The PW$\pi$ machine searches for a matching prefix by non-deterministically making a choice whenever it has to branch at alternation or Kleene nodes. While this machine is not very useful in practice, it allows us to arrive at a precise model for backtracking regular expression matchers. Backtracking matchers operate by attempting all the possible search paths in order; this allows us to model them with a stack of PW$\pi$ machines. We call the resulting machine the PWF$\pi$ machine:
\begin{definition}
The PWF$\pi$ machine consists of a stack of PW$\pi$ machines. The transitions of the PWF$\pi$ machine are given below:
\[
\infer{\pwpi{p}{w} \step \pwpi{q}{w'}}{\pwfpi{p}{w}{f} \step \pwfpi{q}{w'}{f}}
\hspace{3em}
\infer{\pwpi{p}{w} \not\step}{\pwfpi{p}{w}{f} \step f}
\]
\[
\infer{\pwpi{p}{w} \step \pwpi{q_1}{w} \qquad \pwpi{p}{w} \step \pwpi{q_2}{w}}{\pwfpi{p}{w}{f} \step \pwfpi{q_1}{w}{\pwfpi{q_2}{w}{f}}}
\]

The initial state of the PWF$\pi$ machine is $[\pwpi{p_0}{w}]$. The machine may terminate if one of the PW$\pi$ machines locates a match or if none of them succeeds in finding a match. In the latter case the PWF$\pi$ machine has exhausted the entire search space and determined that the input string cannot be matched by the regular expression in question.
\end{definition}
The PWF$\pi$ machine allows us to analyze backtracking regular expression matchers at an abstract level without concerning ourselves about any implementation specific details. More importantly, it gives an accurate cost model of backtracking matchers; the number of steps executed by the PWF$\pi$ machine corresponds to the amount of work a backtracking matcher has to perform when searching for a match. In the following sections we employ these ideas to develop and implement our static analysis.

\section{Static analysis for exponential blowup}
\label{staticanalysis}

The problem we are aiming to solve is this: given a regular expression $e$, represented as in Section~\ref{secmachine}, are  there input strings $x$, $y$, and $z$, such that:
\begin{enumerate}
\item Reading $x$ takes the machine to a pointer $p_0$ that is the root of a Kleene star expression. 
\item Reading the input $w$ takes the machine from $p_0$ back to $p_0$, and in at least two different ways, that is, along two different paths in the NFA.
\item Reading the input $z$ when starting from $p_0$ causes the match to fail.
\end{enumerate}

We call $x$ the \emph{prefix}, $w$ the \emph{pumpable} string by analogy with pumping lemmas in automata theory~\cite{hopcroftullman}, and $z$ the \emph{failure suffix}. 

\begin{figure}[t]
\begin{center}
\begin{tikzpicture}[level distance=4em]
\newcommand\failpath{ edge from parent[->] node[left] {$z$} }
\newcommand\ffail{\textrm{fail}}
\node {$r$}
  child {
  node {$p_0$}
    child {
      node {$p_0$}
    child {
      node {$p_0$}
        child {node {$\ffail$} \failpath } 
      edge from parent[->] node[left] {$w$}
    }
  child {
      node {$p_0$}
        child {node {$\ffail$} \failpath } 
      edge from parent[->] node[left] {$w$}
    }
      edge from parent[->] node[left] {$w$}
    }     
   child[missing] { }       
  child {
      node {$p_0$}
    child {
      node {$p_0$}
        child {node {$\ffail$} \failpath } 
      edge from parent[->] node[left] {$w$}
    }
  child {
      node {$p_0$}
        child {node {$\ffail$} \failpath } 
      edge from parent[->] node[left] {$w$}
    }
      edge from parent[->] node[left] {$w$}
    }
        edge from parent[->] node[left] {$x$}
  } 
;
\end{tikzpicture}
\end{center}
\caption{The search tree for $x\,w\,w\,y$}
\label{figsearchtree}
\end{figure}
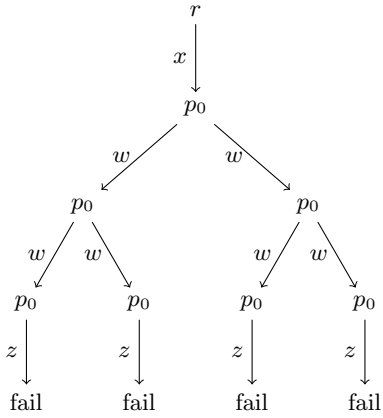

From these three strings, malicious inputs can be constructed: the $n$-th malicious input is $x\,w^n\,z$. Figure~\ref{figsearchtree} illustrates the search tree that a backtracking matcher has to explore when $w$ is pumped twice. Because $w$ can be matched in two different ways, the tree branches every time a $w$ is read from the input. All branches fail in the end due to the trailing $z$, so that the matcher must explore the whole tree.

To state the analysis more formally, we will need to define paths in the matcher.
\begin{definition}
A path of pointers, $t : p \astep{w} q$ is defined according to the following inductive rules:
\begin{itemize}
\item
For each pointer $p$, $[p] : p \astep{\empstring} p$ is a path (identity).
\item
If $t : p \astep{w} q$ is a path and there exists a PW$\pi$ transition such that:
\[
\pwpi{q}{w'w_1} \step \pwpi{q'}{w_1}
\]
Then $\conc{t}{[q']} : p \astep{ww'} q'$ is also a path.
\end{itemize}
\end{definition}

\begin{lemma}\label{lem_wppi_path}
The path $t : p \astep{w} q$ ($q \ne p$) exists if and only if a PW$\pi$ run exists such that:
\[
\pwpi{p}{ww'} \steps \pwpi{q}{w'}
\]
\end{lemma}

Lemma~\ref{lem_wppi_path} associates a unique string $w$ with each path of pointers (the sub-string matched by the corresponding PW$\pi$ run). However, note that the inverse of this implication does not hold; there can be input strings for which we may find more than one PW$\pi$ run. In fact, it is this property of paths that leads us to the main theorem of this paper:
\begin{theorem}\label{exptheorem}
For a given Kleene expression $p_0$ where $\pi(p_0) = \kleene{p_1}$, if at least two paths exist such that $t_1 : p_1 \astep{w} p_0$, $t_2 : p_1 \astep{w} p_0$ and $t_1 \ne t_2$, then a regular expression involving $p_0$ exhibits $o(2^n)$ runtime on a  backtracking regular expression matcher for input strings of the form $xw^nz$ where $x$ is a sub-string matching the \textit{prefix} of $p_0$ and $z$ is such that $xw^nz$ fails to match the overall expression.
\end{theorem}

While a formal proof of Theorem~\ref{exptheorem} is outside of the scope of this paper, we sketch its proof with reference to Figure~\ref{figsearchtree}. The prefix $x$ causes the PWF$\pi$ machine to advance into a state where it has to match $p_0$ against the remainder of the input string, which leads to the branching of the search tree. Finally, the suffix $z$ at the end of the input causes each search path to fail, which in turns forces the PWF$\pi$ machine to backtrack and explore the entire search tree before concluding that a match cannot be found. For the complexity, note that each additional pumping increases the size of the input by a constant (the length of $w$) whereas it doubles the size of the binary subtree given by the $w$ branches, as well as the number of failed attempts to match $z$ at the end. If there are more than 2 ways to match the pumpable string, say $b$, then $b$ rather than $2$ becomes the base of the exponent, but 2 is still a lower bound. The matching of the prefix $x$ at the beginning contributes a constant to the runtime, which can be disregarded relative to the exponential growth. Thus the lower bound for the number of steps is exponential.

\newcommand\wpm[2]{(#1, #2)}
\newcommand\set[1]{\{#1\}}
\newcommand\flist[2]{\mathcal{F}(#1, #2)}
\newcommand\dv[2]{\mathcal{D}_{#1}(#2)}
\subsection{Generating the pumpable string}
The most important step in generating an attack string for a vulnerable regular expression is to generate the pumpable string $w$ in $xw^nz$ (for some Kleene sub-expression). 
In order to arrive at the machine for building the pumpable string, we must first introduce several utility definitions. Note that in the remainder of this discussion, $p_0$ refers to a Kleene expression such that $\pi(p_0) = \kleene{p_1}$.

\begin{definition}
For a given pointer $p$, the operation $\evolve{p}$ (called \textit{evolve}) is defined as:
\[
\evolve{p} = [q \mid \exists t. t : p \astep{\empstring} q \land \exists a. \pi(q) = a]
\]
Notice that the result of $\evolve{p}$ is a list of pointers.
\end{definition}

\begin{definition}
The function $\dv{a}{P}$, (called \textit{derive}) is defined on a list of pointers $P$ and an input symbol $a$ 
 according to the following rules:
\begin{eqnarray*}
\dv{a}{\emplist} &=& \emplist\\
\dv{a}{\cons{h}{t}} &=&
\begin{cases}
\dv{a}{t} & \text{ if } \pi(h) = b, b \ne a\\
\cons{q}{\dv{a}{t}} & \text{ if } \pi(h) = a \land \knode{h} = q\\
\dv{a}{\conc{\evolve{h}}{t}} & \text{ otherwise.}
\end{cases}
\end{eqnarray*}
\end{definition}

The definition $\dv{a}{P}$ is analogous to Brzozowski's derivatives of regular expressions~\cite{1964_brzozowski}. 
 In essence, the analysis computes derivatives of a Kleene expression in order to find two different matcher states for the same input string.
 
\begin{definition}
A wP frame is defined as a pair $(w, P)$ where $w$ is a string and $P$ is a list of pointers. A non-deterministic transition relation is defined on wP frames as follows:
\[
\infer{\dv{a}{P} \ne \emplist}{(w,P) \rightarrow (w\,a, \dv{a}{P})}
\]
\end{definition}

\newcommand\hfpi[2]{\mconf{\mpair{#1}{#2}}}
\begin{definition}
The HF$\pi$ machine has configurations of the following form:
\[
\hfpi{H}{f}
\]
Here $H$ (history) represents a set of (sorted) pointer lists and $f$ is a list of wP frames. A deterministic transition relation defines the behavior of this machine as follows:
\[
\infer{(w, P) \step (wx_0, P_0) \quad \ldots \quad (w, P) \step (wx_n, P_n) \quad \forall i. x_i \in \Sigma \quad P_i \notin H}{\hfpi{H}{\cons{(w, P)}{f}} \step \hfpi{H \cup \set{P_0, \ldots, P_n}}{\conc{f}{[(wx_0, P_0), \ldots, (wx_n, P_n)]}}}
\]
The initial configuration of the HF$\pi$ machine is $\hfpi{\empset}{[(\empstring, [p_1])]}$ and the machine can terminate in either of the following two configurations:
\[
\hfpi{H}{\emplist}
\]
\[
\hfpi{H}{\cons{(w, P)}{f}} \text{ where } \exists p', p'' \in P.\;\exists t', t''.\;t' : p' \astep{\empstring} p_0 \land t'' : p'' \astep{\empstring} p_0
\]
In the former configuration the machine has determined the Kleene expression in question to be non-vulnerable while in the latter it has derived the pumpable string $w$.
\end{definition}

\subsection{Generating the Prefix and the Suffix}

For a regular expression of the form $e_1\,(\kleene{e_2})\,e_3$, apart from a pumpable string $w$, we must also generate a prefix $x$ and a suffix $z$. The intention is that $x$ would lead the matcher to the point where it has to match $\kleene{e_2}$, after which we can pump many copies of $w$ to increase the search space of the matcher. However, a successful exploit also needs a suffix $z$ which forces the matcher to fail and so to traverse the entire search space.

Generating the prefix is quite straightforward since a depth-first search for Kleene sub-expressions (on the AST) can be augmented such that a (minimal) prefix is generated for each search-path. On the other hand, the suffix generation is more involved. One would think $z$ should be generated such that it fails to match the continuation expression $e_3$ of the original expression, but this intuition is flawed since there is a possibility that $z$ could be matched by $e_2$ itself (while it was meant for $e_3$). Depending on $e_2$, it could be the case that no failure suffix exists. One example we found is that $e_2$ ends in \texttt{.*}, so that it can match anything. In other cases, a failure suffix may exists, but depend in complicated ways on $e_2$.
We chose not to solve this problem in full generality, but rather to employ heuristics that find failure suffixes   
 for many practical expressions, as illustrated in the results section.

\section{Implementation of the static analysis}
\label{implementation}

We implemented the HF$\pi$ machine described in Section~\ref{staticanalysis} using the OCaml programming language. OCaml is well suited to programming abstract syntax, and hence a popular choice for writing static analyses.
One of the major obstacles faced with the implementation is that in order to be able to analyze real-world regular expressions, it was necessary to build a sophisticated parser. In this regard, we decided to support the most common elements of the Perl / PCRE standards, as these seem to be the most commonly used (and adapted) syntaxes. It should be noted that the current implementation does not support back-references or look-around expressions due to their inherent complexity; it remains to be seen if the static analysis proposed in this work can be adapted to handle such ``regexes". However, as it was explained earlier, exponential vulnerabilities in pattern specifications are not necessarily dependent on the use of back-references or other advanced constructs (although one would expect such constructs to further increase the search space of a backtracking matcher). A detailed description of the pattern specification syntax currently supported by the implementation has been included in the resources accompanying this paper.

The implementation closely follows the description of the HF$\pi$ machine presented in Section~\ref{staticanalysis}. The history component $H$ is implemented as a set of sorted integer lists, where a single sorted integer list corresponds to a list of nodes pointed by the pointer list $P$ of a wP frame $(w, P)$. This representation allows for quick elimination of looping wP frames. While the size of $H$ is potentially exponential in the number of nodes of a given Kleene expression, for practical regular expressions we found this size to be well within manageable levels (as evidenced in the results section).

A matter not addressed in the current work is that the PWF$\pi$ machine (and  naive backtracking matching algorithms in general) can enter into infinite loops for Kleene expressions when the enclosed sub-expression matches the empty string (i.e the sub-expression is nullable). Although a complete treatment of this issue and its solution (implemented by most of the well known backtracking matchers) is beyond the scope of this paper, it should be mentioned that a similar problem occurs in the HF$\pi$ machine during the $\evolve{p}$ operation. We have incorporated a method for detecting and terminating such infinite loops into the OCaml code for the $\evolve{p}$ function so that it terminates in all cases.

\section{Experimental results}
\label{results}

The analysis was tested on two corpora of regexes (Figure~\ref{expresults}). The first of these was extracted from an online regex library called \textit{RegExLib}~\cite{2012_regexlib}, which is a community-maintained regex archive; programmers from various disciplines submit their solutions to various pattern matching tasks, so that other developers can reuse these expressions for their own pattern matching needs. The second corpus was extracted from the popular intrusion detection and prevention system \textit{Snort}~\cite{2012_snort}, which contains regex-based pattern matching rules for inspecting IP packets across network boundaries. The contrasting purposes of these two corpora allow us to get a better view of the seriousness of exponential vulnerabilities in practical regular expressions.

\begin{table}[t]
\begin{center}
\def\arraystretch{1.2}
\setlength{\tabcolsep}{1ex}
\begin{tabular}{|l||r|r|}
\hline
& RegExLib & Snort\\
\hline
\hline
Total patterns & 2994 & 12499\\
\hline
Analyzable (only regular constructs) & 2213 & 9408\\
\hline
Uses Kleene star & 1103 & 2741\\
\hline
Pumpable Kleene and suffix found & 127 & 15\\
\hline
Pumpable Kleene only & 20 & 4\\
\hline
No pumpable Kleene  & 2066 & 9389\\
\hline
Max HF$\pi$ steps & 509 & 256 \\
\hline
Total classification time   & 40 s& 10 s 
\\
(Intel Core 2 Duo 1.8 MHz, 4 GB RAM) &&
\\
\hline
\end{tabular}
\end{center}
\caption{Experimental results with RegExLib and Snort}
\label{expresults}
\end{table}

The regex archive for RegExLib was only available through the corresponding website~\cite{2012_regexlib}. Therefore, as the first step the expressions had to be scraped from their web source and adapted so that they can be fed into our tool. These adaptations include removing unnecessary white-space, comments and spurious line breaks. A detailed description of these adjustments as well as copies of both adjusted and un-adjusted data sets have been included with the resources accompanying this paper (also including the Python script used for scraping). The regexes for Snort, on the other hand, are embedded within plain text files that define the Snort rule set. A Python script (also included in the accompanying resources) allowed the extraction of these regexes, and no further processing was necessary. 

The results of the HF$\pi$ static analysis on these two corpora of regexes are presented in Table~\ref{expresults}.
The figures  show that we can process  around 75\% of each of the corpora with the current level of syntax support. Out of these analyzable amounts, it is notable that regular expressions from the RegExLib archive use the Kleene operator more frequently (about 50\% of the analyzable expressions) than those from the Snort rule set (close to 30\%). About 11.5\% of the Kleene-based RegExLib expressions were found to have a pumpable Kleene expression as well as a suitable suffix, whereas for Snort this figure stands around 0.55\%.

The vulnerabilities reported range from trivial programming errors to more complicated cases. For an example, the following regular expression is meant to validate time values in 24-hour format (from RegExLib):
\begin{verbatim}
    ^(([01][0-9]|[012][0-3]):([0-5][0-9]))*$
\end{verbatim}
Here the author has mistakenly used the Kleene operator instead of the \texttt{?} operator to suggest the presence or non-presence of the value. This pattern works perfectly for all intended inputs. However, our analysis reports that this expression is vulnerable with the pumpable string ``\verb|13:59|" and the suffix ``\verb|/|". This result gives the programmer a warning that the regular expression presents a DoS security risk if exposed to user-malleable input strings to match.

 For a moderately complicated example,  consider the following regular expression (again from RegExLib):
\begin{verbatim}
^([a-zA-z]:((\\([-*\.*\w+\s+\d+]+)|(\w+)\\)+)(\w+.zip)|(\w+.ZIP))$
\end{verbatim}
This expression is meant to validate file paths to zip archives. Our tool identifies this expression as vulnerable and generates the prefix ``\verb|z:\ |", the pumpable string ``\verb|\zzz\|" and the empty string as the suffix. This is probably an unexpected input in the author's eye, and this is another way in which our tool can be useful in that it can point out potential mis-interpretations which may have materialized as vulnerabilities. 

It is worth noting that the HF$\pi$ machine manages to classify both the corpora (the analyzable portions) in a matter of seconds on modest hardware. This shows that our static analysis is usable for most practical purposes, with the average classification time for an expression in the range of micro-seconds. The two extreme cases for which the machine took several seconds for the classification are given below (only the respective Kleene expressions):
\begin{verbatim}
    ([\d\w][-\d\w]{0,253}[\d\w]\.)+

    ([^\x00]{0,255}\x00)*
\end{verbatim}

Here  counting expressions 
\verb|[-\d\w]{0,253}| and \verb|[^\x00]{0,255}| 
were expanded out during the parsing phase. The expansion produces a large Kleene expression, which naturally requires more analysis during the HF$\pi$ simulation. However, it should be noted that such expressions are the exception rather than the norm.

Finally, it should be mentioned that all the vulnerabilities reported above were individually verified using a modified version of the PWF$\pi$ machine (which counts the number of steps taken for a particular matching operation). A sample of those vulnerabilities was also tested on the Java regular expression matcher. 

\section{Conclusions}
\label{conclusions}

We have presented a static analysis to help programmers defend against regular expression DoS attacks. Large numbers of regular expressions can be analysed quickly, and developers are given feedback on where in their regular expressions the problem has been identified as well as examples of malicious input.

As illustrated in Section~\ref{results}, the prefix, pumpable string and failure suffix can be quite short. If their length is, say, 3, 5 and 0 characters, then an attacker only needs to spend a very small amount of effort in providing a malicious input of length 3+5*100 characters to cause a matching time in excess of $2^{100}$ steps. Even if a matching step takes only a nanosecond, such a running time takes, for all intents and purposes, forever. The attacker can still scale up the attack by pumping a few times more and thereby correspondingly multiplying the matching time.

The fact that  the complexity of checking a regular expression for exponential runtime may be computationally expensive in the worst case does not necessarily imply that such an analysis is futile. Type checking in functional languages like ML and Haskell also has high complexity~\cite{mairson1989deciding,seidl1994haskell}, yet works efficiently in practice because the worst cases rarely occur in real-world code. There are even program analyses for undecidable problems like termination~\cite{berdine2006automatic}, so that the worst-case running time is infinite; what matters is that the analysis produces results in enough cases to be useful in practice. It is a common situation in program analysis that tools are not infallible (having false positives and negatives), but they are nonetheless useful for identifying points in code that need attention by a human expert~\cite{dowd-software}.  
\subsection{Related work}

A general class of DoS attacks based on algorithmic complexities has been explored in~\cite{2003_dos_crosby}. In particular, the exponential runtime behavior of backtracking regular expression matchers has been discussed in~\cite{2007_regex_cox} and~\cite{2012_redos_checkmarx}. The seriousness of this issue is further expounded in~\cite{2006_btrack_smith} and~\cite{2012_ids_namjoshi} where the authors demonstrate the mounting of DoS attacks on an IDS/IPS system (Snort) by exploiting the said vulnerability. The solutions proposed in these two works involve modifying the regular expressions and/or the matching algorithm in order to circumvent the problem in the context of IDS/IPS systems. We consider our work to be quite orthogonal and more general since it is based on a compile-time static analysis of regular expressions. However, it should be noted that both of those works concern of regexes with back-references, which is a feature we are yet to explore (known to be NP-hard~\cite{1990_aho}).


While the problem of ReDoS has been known for at least a decade, we are not aware of any previous static analysis for defending against it. A  handful of tools exist that can assist programmers in finding such vulnerable regexes. Among these tools we found  Microsoft's SDL Regex Fuzzer~\cite{2011_regex_fuzzer} and the RegexBuddy~\cite{2012_regexbuddy} to be the most usable implementations, as other tools were too unstable to be tested with complex expressions. 

While RegexBuddy itself is not a security oriented software, it offers a debug mode, which can be used to detect what the authors of the tool refer to as \textit{Catastrophic Backtracking}~\cite{2009_catas_goyvaerts}. Even though such visual debugging methods can assist in detecting potential vulnerabilities, it would only be effective if the attack string is known in advance---this is where a static analysis method like the one presented on this paper has a clear advantage. 

SDL Fuzzer, on the other hand, is aimed specifically at analyzing regular expression vulnerabilities. While details of the tool's internal workings are not publicly available, analyzing the associated documentation reveals that it operates fuzzing, i.e., by brute-forcing a sequence of generated strings through the regular expression in question to detect long running times. The main disadvantage of this tool is that it can take a very long time for the tool to classify a given expression. Tests using some of the regular expressions used in the results section above revealed that it can take up to four minutes for the Fuzzer to classify certain expressions. It is an inherent limitation of fuzzers for exponential runtime DoS attacks that the finding out if something takes a long time by running it takes a long time. By contrast, our analysis statically analyzes an expression without ever running it. It is capable of classifying thousands of regular expressions in a matter of seconds. Furthermore, the output produced by the SDL Fuzzer only reports the fact that the expression in question failed to execute within a given time limit for some input string. Using this generated input string to pin-point the exact problem in the expression would be quite a daunting task. In contrast, our static analysis pin-points the exact Kleene expression that causes the vulnerability and allows programmers to test their matchers with a sequence of malicious inputs.

\subsection{Directions for further research}

In further work, we aim to broaden the coverage of our tool to include more regexes. Given its basis in our earlier work on abstract machines~\cite{2011_regexpsos} and derivatives~\cite{1964_brzozowski}, we aim for a formal proof of the correctness of our analysis. 
We intend to release the source code of the tool as an open source project. More broadly, we hope that raising awareness of the dangers of backtracking matchers will help in the adoption of superior techniques for regular expression matching~\cite{2009_regex_cox,1968_thompson,2011_regexpsos}.

\bibliographystyle{plain}
\bibliography{reg-exp-sec}

\begin{thebibliography}{10}

\bibitem{1990_aho}
Alfred~V. Aho.
\newblock Algorithms for {F}inding {P}atterns in {S}trings.
\newblock In Jan van Leeuwen, editor, {\em Handbook of theoretical computer
  science (vol. A)}, pages 255--300. MIT Press, Cambridge, MA, USA, 1990.

\bibitem{2007_dragonbook}
Alfred~V. Aho, Monica Lam, Ravi Sethi, and Jeffrey~D. Ullman.
\newblock {\em Compilers - Principles, Techniques and Tools}.
\newblock Addison Wesley, second edition, 2007.

\bibitem{berdine2006automatic}
J.~Berdine, B.~Cook, D.~Distefano, and P.~OÕHearn.
\newblock Automatic termination proofs for programs with shape-shifting heaps.
\newblock In {\em Computer Aided Verification}, pages 386--400. Springer, 2006.

\bibitem{1964_brzozowski}
Janusz~A. Brzozowski.
\newblock {D}erivatives of {R}egular {E}xpressions.
\newblock {\em J. ACM}, 11(4):481--494, 1964.

\bibitem{chess2004static}
B.~Chess and G.~McGraw.
\newblock Static analysis for security.
\newblock {\em Security \& Privacy, IEEE}, 2(6):76--79, 2004.

\bibitem{2007_regex_cox}
Russ Cox.
\newblock {R}egular {E}xpression {M}atching {C}an {B}e {S}imple {A}nd {F}ast
  (but is slow in {J}ava, {P}erl, {P}hp, {P}ython, {R}uby, ...).
\newblock Available at \url{http://swtch.com/\textasciitilde
  rsc/regexp/regexp1.html}, January 2007.

\bibitem{2009_regex_cox}
Russ Cox.
\newblock Regular expression matching: the virtual machine approach.
\newblock Available at \url{http://swtch.com/\textasciitilde
  rsc/regexp/regexp2.html}, December 2009.

\bibitem{2003_dos_crosby}
Scott~A. Crosby and Dan~S. Wallach.
\newblock {D}enial of {S}ervice via {A}lgorithmic {C}omplexity {A}ttacks.
\newblock In {\em Proceedings of the 12th USENIX Security Symposium},
  Washington, DC, August 2003.

\bibitem{dowd-software}
Mark Dowd, John McDonald, and Justin Schuh.
\newblock {\em The Art of Software Security Assessment: Identifying and
  Preventing Software Vulnerabilities}.
\newblock Addison Wesley, 2006.

\bibitem{2009_catas_goyvaerts}
Jan Goyvaerts.
\newblock {R}unaway {R}egular {E}xpressions: {C}atastrophic {B}acktracking.
\newblock Available at
  \url{http://www.regular-expressions.info/catastrophic.html}, 2009.

\bibitem{hopcroftullman}
John~E. Hopcroft and Jeffrey~D. Ullman.
\newblock {\em Introduction to Automata Theory, Languages and Computation}.
\newblock Addison-Wesley, 1979.

\bibitem{livshits2005finding}
V.B. Livshits and M.S. Lam.
\newblock Finding security vulnerabilities in java applications with static
  analysis.
\newblock In {\em Proceedings of the 14th conference on USENIX Security
  Symposium}, volume~14, pages 18--18, 2005.

\bibitem{2012_regexbuddy}
Just Great Software~Co. Ltd.
\newblock {R}egex{B}uddy.
\newblock Available at \url{http://www.regexbuddy.com/}, 2012.

\bibitem{mairson1989deciding}
H.G. Mairson.
\newblock Deciding {ML} typability is complete for deterministic exponential
  time.
\newblock In {\em Proceedings of the 17th {ACM} {SIGPLAN-SIGACT} symposium on
  Principles of programming languages}, pages 382--401. ACM, 1989.

\bibitem{2011_regex_fuzzer}
Microsoft.
\newblock {SDL} {R}egex {F}uzzer.
\newblock Available at
  \url{http://www.microsoft.com/en-gb/download/details.aspx?id=20095}, 2011.

\bibitem{2012_ids_namjoshi}
Kedar Namjoshi and Girija Narlikar.
\newblock {R}obust and {F}ast {P}attern {M}atching for {I}ntrusion {D}etection.
\newblock In {\em Proceedings of the 29th conference on Information
  communications}, INFOCOM'10, pages 740--748, Piscataway, NJ, USA, 2010. IEEE
  Press.

\bibitem{2012_redos_owasp}
The Open Web Application Security~Project (OWASP).
\newblock {R}egular {E}xpression {D}enial of {S}ervice - {ReDoS}.
\newblock Available at
  \url{https://www.owasp.org/index.php/Regular_expression_Denial_of_Service_-_%
ReDoS}, 2012.

\bibitem{2011_regexpsos}
Asiri Rathnayake and Hayo Thielecke.
\newblock {R}egular {E}xpression {M}atching and {O}perational {S}emantics.
\newblock In {\em Structural Operational Semantics (SOS 2011)}, Electronic
  Proceedings in Theoretical Computer Science, 2011.

\bibitem{2012_regexlib}
RegExLib.com.
\newblock {R}egular {E}xpression {L}ibrary.
\newblock Available at \url{http://regexlib.com/}, 2012.

\bibitem{2012_redos_checkmarx}
Alex Roichman and Adar Weidman.
\newblock {R}egular {E}xpression {D}enial of {S}ervice.
\newblock Available at
  \url{http://www.checkmarx.com/white_papers/redos-regular-expression-denial-o%
f-service/}, 2012.

\bibitem{seidl1994haskell}
H.~Seidl et~al.
\newblock Haskell overloading is {DEXPTIME}-complete.
\newblock {\em Information Processing Letters}, 52(2):57--60, 1994.

\bibitem{2006_btrack_smith}
Randy Smith, Cristian Estan, and Somesh Jha.
\newblock {B}acktracking {A}lgorithmic {C}omplexity {A}ttacks {A}gainst a
  {NIDS}.
\newblock In {\em Proceedings of the 22nd Annual Computer Security Applications
  Conference}, ACSAC '06, pages 89--98, Washington, DC, USA, 2006. IEEE
  Computer Society.

\bibitem{2012_snort}
Sourcefire.
\newblock {S}nort {(IDS/IPS)}.
\newblock Available at \url{http://www.snort.org/}, 2012.

\bibitem{1968_thompson}
Ken Thompson.
\newblock {P}rogramming {T}echniques: {R}egular {E}xpression {S}earch
  {A}lgorithm.
\newblock {\em Communications of the ACM}, 11(6):419--422, June 1968.

\end{thebibliography}

\end{document}